\documentclass[11pt, a4paper, prd, showpacs, twocolumn, preprintnumbers]{revtex4}

\usepackage{psfig, epsfig,color,amsmath,amssymb, mathrsfs}
\usepackage{graphicx}
\usepackage{fancyhdr, url, amsfonts}
\newcommand{\be}[1]{\small \begin{equation} #1 \end{equation}}

\newcommand{\abs}[1]{\left| #1 \right|}

\newcommand{\bparen}[1]{\left[ #1 \right]}
\newcommand{\paren}[1]{\left( #1 \right)}

\newcommand{\goodspace}{\hspace{0.6cm}}
\newcommand{\onehalf}{\frac{1}{2}}
\newcommand{\dis}{\displaystyle}

\def\apriori{{\it a priori\/}}
\def\viceversa{{\it vice versa\/}}
\def\eg{{\it e.g.\ }}

\def\ie{{\it i.e.\ }}

\begin{document}

\title{Geometry of Higher-Dimensional Black Hole Thermodynamics}

\author{Jan E. \AA man}
\email{ja@physto.se}

%\author{Ingemar Bengtsson}
%\email{ingemar@physto.se}

\author{Narit Pidokrajt}
\email{narit@physto.se}

\affiliation{Quantum and Field Theory group, Department of Physics, AlbaNova \\ 
Stockholm University,  SE-106 91 Stockholm, Sweden.}
\homepage{www.kof.physto.se}

\date{\today}

\begin{abstract}

We investigate thermodynamic curvatures of the Kerr and Reissner-Nordstr\"om (RN) black holes in spacetime dimensions higher than four. These black holes possess thermodynamic geometries similar to those in four dimensional spacetime. The thermodynamic geometries are the Ruppeiner geometry and the conformally related Weinhold geometry. The Ruppeiner geometry for $d=5$ Kerr black hole is curved and divergent in the extremal limit. For $d \geq 6$ Kerr black hole there is no extremality but the Ruppeiner curvature diverges where one suspects that the black hole becomes unstable.  The Weinhold geometry of the Kerr black hole in arbitrary dimension is a flat geometry.  For RN black hole the Ruppeiner geometry is flat in all spacetime dimensions, whereas its Weinhold geometry is curved. In $d \geq 5$ the Kerr black hole can possess more than one angular momentum. Finally we discuss the Ruppeiner geometry for the Kerr black hole in $d=5$ with double angular momenta. 
\end{abstract}


\preprint{Stockholm}

\preprint{USITP 05-05} 

%\preprint{hep-th/0510139}

\maketitle

\section{Introduction}

The Hessian matrix of the thermodynamic entropy is known as the Ruppeiner metric \cite{ruppeiner}. It is a metric defined on the state space by 
\be{
g^{R}_{ij} = - \partial_i \partial_j S(M, N^a),
}
where $S$ is the entropy, $M$ denotes the energy and $N^a$ are other extensive variables of the system. There have been a number of results indicating that this geometry measures the underlying statistical mechanics of the system. In particular for systems with no statistical mechanical interactions (\eg the ideal gas), the Ruppeiner geometry is flat and \viceversa. Furthermore it appears that a divergent Ruppeiner curvature indicates a phase transition \cite{ruppeiner, idealgas, johnston}. There is another metric which is defined as the Hessian of the energy (mass), it is known as the Weinhold metric \cite{weinhold, gibbons} 
\be{
g^W_{ij} = \partial_i \partial_j M(S, N^a).
}
The Ruppeiner and Weinhold metrics are related to each other \cite{conformal1, conformal2} via
\be{
\label{eq:conformal}
ds^2_R = \frac{1}{T} ds^2_W.
}
Since black holes are regarded as thermodynamic systems \cite{hawking}, it is then natural to investigate their thermodynamic geometries. Previous studies \cite{ourpaper, china} suggest that the thermodynamic geometry of the black holes has a pattern from which one may possibly deduce physical insights. In \cite{ourpaper} it is found that the divergence of the Ruppeiner curvature of the Kerr black hold indicates a phase transition.

As a generalization of our previous work \cite{ourpaper}, we will apply the Ruppeiner thermodynamic theory to the higher dimensional black hole solutions which were first derived by Myers and Perry in 1986 \cite{myers-perry}, in hopes to obtain further structure of black holes and the Ruppeiner theory itself. As it turns out, we obtain interesting results which may be a justification for a possible application of the Ruppeiner theory to black hole solutions that exist in various gravity theories \eg string theory.   

\section{Reissner-Nordstr\"om black hole}

The Reissner-Nordstr\"om black hole is a solution of the Einstein equation coupled to the Maxwell field. In arbitrary spacetime dimension it is given by 
\be{
ds^2 = -V dt^2 + V^{-1} dr^2 + r^2 d\Omega^2_{(d-2)}
}
where $d\Omega^2_{(d-2)}$ is the line element on the $(d-2)$ unit sphere with $d$ being a spacetime dimension. The volume of the $(d-2)$ unit sphere is given by 
\be{
\Omega_{(d-2)} = \frac{2\pi^{\frac{d-1}{2}}}{\Gamma(\frac{d-1}{2})}.
}
$V$ is a function of mass and charge given in terms of parameters $\mu$ and $q$
\be{
V = 1 - \frac{\mu}{r^{d-3}} + \frac{q^2}{r^{2(d-3)}}
}
where 
\be{
\mu = \frac{16\pi G M}{(d-2) \Omega_{(d-2)}}
}
and
\be{
q = \sqrt{\frac{8\pi G}{(d-2)(d-3)}} \:Q.
}
An event horizon of the RN black hole is where $V=0$, which can be solved analytically in arbitrary dimension. For the sake of tidiness and simplicity, we set Newton's gravitational constant to be $G = \Omega^2_{(d-2)}/16\pi$ in order to eliminate all the $\pi$'s under the square root in (\ref{eq:RNhorizon}). There are two roots, one of which is an outer horizon, $r_{+}$ while the other is called a Cauchy horizon
\be{
\label{eq:RNhorizon}
r_{\pm} = \paren{ \frac{\mu}{2} \pm \frac{\mu}{2}\sqrt{1 - \frac{4q^2}{\mu^2}} \: }^{1/(d-3)}.
}
It is obviously seen that 
\be{
r^{d-3}_{+}  + r^{d-3}_{-} = \mu \goodspace \text{and} \goodspace r^{d-3}_{+} r^{d-3}_{-} = q^2.
} 
This solution develops a singularity when $\dis q^2 > \mu^2 / 4$ with the singularity at $r=0$. When $\dis q^2 < \mu^2 / 4$, we have the outer event horizon as in (\ref{eq:RNhorizon}). We note that $\mu$ and $q$ are the ADM mass and electric charge of the black hole respectively \cite{myers-perry}. When expressed in terms of the mass, charge and dimensionality of the RN black hole, we obtain
\be{
r^{d-3}_{+} = \frac{M\Omega_{(d-2)}}{2(d-2)}\paren{1 + \sqrt{1 - \frac{d-2}{2(d-3)}\frac{Q^2}{M^2}} \,}.
}
In arbitrary dimension the black hole becomes extremal when
\be{
\frac{Q^2}{M^2} = \frac{2(d-3)}{d-2}.
}
The area of the event horizon of the RN black hole is thus given by 
\be{
A = \Omega_{(d-2)} r^{(d-2)}_{+}.
}
The entropy of the hole \cite{hehl} takes the form 
\be{
S = \frac{k_B A }{4 G \hbar} = \frac{k_B}{4G} \Omega_{(d-2)} r^{(d-2)}_+ ,  
}
with $\hbar = 1$ for simplicity. We can further introduce Boltzmann's constant to absorb
the $\pi$'s and other numbers in the following way,
\be{
k_B = \frac{[2(d-2)]^{\frac{d-2}{d-3}}}{4\pi \Omega^{\frac{1}{d-3}}_{(d-2)} } .
}
Hence we obtain the entropy function as
\be{
\dis
S = r^{(d-2)}_+ = \paren{r^{d-3}_+}^{\frac{d-2}{d-3}}. 
}
Explicitly in terms of mass and charge, it reads
\be{
\label{eq:RNentropy}
S = \paren{M + M\sqrt{1 - \frac{d-2}{2(d-3)}\frac{Q^2}{M^2}}\, }^{\dis \frac{d-2}{d-3}}.
}
We have learned from our previous work \cite{ourpaper} that for the RN black hole, it is simpler to work in Weinhold coordinates. An inversion of (\ref{eq:RNentropy}) gives $M$ as 
\be{
\dis 
%M = \frac{1}{2S^{\paren{\frac{d-3}{d-2}}}}\paren{ S^{\frac{2(d-3)}{d-2}} + \frac{d-2}{2(d-3)}Q^2 }.
M = \frac{S^{\frac{d-3}{d-2}} }{2} + \frac{d-2}{4(d-3)}\frac{Q^2}{S^{\frac{d-3}{d-2}}}.
}
In $d=4$ it takes a simple form
\be{
M = \frac{\sqrt{S}}{2}\paren{1 + \frac{Q^2}{S}}.
}
The Weinhold metric of the RN black hole is diagonalizable in any dimension by choosing the new coordinate
\be{
%u = \frac{Q}{\sqrt{\frac{2(d-3)}{(d-2)}} S^{\frac{d-3}{d-2}}}.
u = \sqrt{\frac{d-2}{2(d-3)}}\frac{Q}{S^{\frac{d-3}{d-2}}}, 
}
where $ -1 \leq u \leq  1$. Thus we obtain the diagonal Weinhold metric as
\be{
ds_W^2 = S^{-\frac{d-1}{d-2}} \bigg[ -\onehalf \frac{d-3}{(d-2)^2}(1- u^2) \: dS^2 +  S^2 \: du^2  \bigg].
}
This metric is curved and Lorentzian. By means of conformal transformation (\ref{eq:conformal}), we obtain the diagonalized Ruppeiner metric for the $d$-dimensional RN black hole in the new coordinates as 
\be{
\label{eq:RN-diag-metric}
ds^2_R = \frac{-dS^2}{(d-2)S} +  \frac{2(d-2)S}{(d-3)}\frac{du^2}{1-u^2},
}
which is a flat metric. The black hole's temperature is given by 
\be{
T = \frac{\partial M}{\partial S} = \frac{d-3}{2(d-2)} \frac{1- u^2}{S^{1/(d-2)}}.
}

%-----------------------------------------------------------
\begin{figure}
\centering
\psfig{file=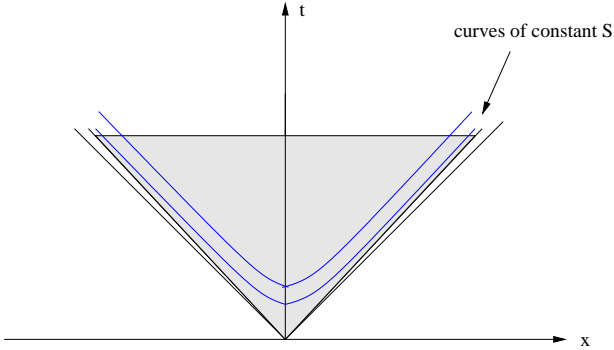, width=.5\textwidth}
\caption{The state space of the four-dimensional Reissner-Nordstr\"om black holes shown as a wedge in a flat Minkowski space. Note that the curves of constant entropy reach the edge the wedge.}
\label{fig:wedge}
\end{figure}
%-----------------------------------------------------------

\noindent Furthermore, we can introduce new coordinates so that the metric in (\ref{eq:RN-diag-metric}) can be written in Rindler coordinates as 
\be{
ds^2 = -d\tau^2 + \tau^2 d\sigma^2,
}
using
\be{
\tau = 2\sqrt{\frac{S}{d-2}} \goodspace \text{and} \goodspace  \sin \frac{\sigma \sqrt{2(d-3)}}{d-2}   = u.
}
It is readily seen that $\sigma$ lies within the following interval
\be{
\label{range}
-{\frac{d-2}{2\sqrt{2(d-3)}}}\pi \leq \sigma \leq  {\frac{d-2}{2\sqrt{2(d-3)}}}\pi.
}
This can be turned into Minkowski coordinates $t$ and $x$ via the following coordinate transformations
\be{
\label{eq:Minkowski-coord}
\begin{split}
t &= \tau \cosh \sigma, \\
x &= \tau \sinh \sigma,
\end{split}
}
such that 
\be{
ds^2 = -dt^2 + dx^2  = -d\tau^2 + \tau^2 d\sigma^2.
%    &= (d\tau \cosh \sigma + \tau \sinh d\sigma )^2 + (d\tau \sinh \sigma + \tau \cosh \sigma)^2 \\      
}
Hence we obtain a Rindler wedge whose opening angle depends on the dimensionality of the RN  black hole, \ie
\be{
\tanh {-\frac{(d-2)\pi}{2\sqrt{2(d-3)}}}  \leq \frac{x}{t} \leq \tanh {\frac{(d-2)\pi}{2\sqrt{2(d-3)}}}.
}
For $d=4$ the resulting wedge is shown in FIG.~\ref{fig:wedge}. It is noticeable that the opening angle of the wedge of the RN black hole grows and reaches the lightcone as $d \rightarrow \infty$. We represent the entropy function of the $d=4$ RN black hole in the Minkowskian coordinates as
\be{
S = \onehalf (t^2 - x^2).
}
\noindent Curves of constant $S$ are segments of hyperbolas.

\section{Kerr black hole}

\begin{figure}
\centering
\psfig{file=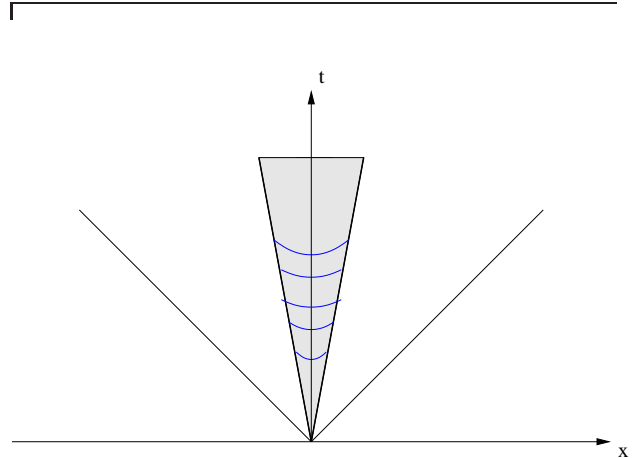, width=.5\textwidth}
\caption{The state space of $d=4$ Kerr black holes shown as a wedge in a flat Minkowski space. The slope of the wedge measures approximately 80$^\circ$ from the $x$-axis. Curves of constant entropy give causal structure to the state space of the black hole.}
\label{fig:wedgeKerr}
\end{figure}

The electrically charged rotating black hole is known as the Kerr-Newman black hole \cite{wheeler, novikov}. The limiting case where the electric charge is zero is known as the Kerr solution. In higher dimensional spacetime there can be more than one angular momentum in the Kerr solution.  For the Kerr black hole with a single nonzero spin \cite{myers-perry, horowitz, ultra-spinning, stojkovic},  we obtain the outermost event horizon by solving the equation
\be{
\label{eq:KerrHorizon}
r^2_{+} + a^2 - \frac{\mu}{r^{d-5}_{+}} = 0.
}
We take our liberty to set the Newton's constant $G = \Omega_{(d-2)}/4\pi$ for the sake of simplification. 
The area of the event horizon is given by
\be{
A = \Omega_{(d-2)} r^{d-4}_+ \: (r^2_+ +  a^2 ).
}
The ADM mass of the hole is defined by  
\be{
\label{eq:ADMmass}
\mu = \frac{4 M}{d-2}.
}
The angular momentum per unit mass is dimension-dependent, namely
\be{
a = \frac{d-2}{2}\frac{J}{M}.
}
By setting $\dis k_B = 1/\pi$ we obtain the entropy function of the Kerr black hole in $d$-dimension as 
\be{
S = r^{d-4}_{+} (r^2_{+} + a^2) = r_{+} \mu.
}
With further algebraic manipulation, we find that even though an explicit entropy function in arbitrary dimension cannot be obtained, we can still work in $d$ dimensions via the Weinhold metric. The mass function in arbitrary $d$ can be written in terms of $S$ and $J$  as 
\be{
\label{eq:Kerr-Mass}
M = \frac{d-2}{4}S^{\frac{d-3}{d-2}} \paren{1 + \frac{4J^2}{S^2}}^{1/(d-2)}.
}
The Weinhold metric $g^W_{ij} = \partial_i \partial_j M(S, J)$ of the $d$-dimensional Kerr black hole takes a complicated form, but it is found to be flat as anticipated \apriori \, based on our previous work \cite{ourpaper}. It takes the form
\begin{widetext}
\be{
\label{eq:Kerr-Ruppeiner}
\begin{split}
ds^2_W &= \lambda\bigg(\bparen{-48(d-5)J^4 + 24S^2 J^2 - (d-3)S^4} dS^2  +  \bparen{64(d-5)J^3 S - 16(d-1)JS^3} dS dJ  \\ 
 &+  \bparen{ -32(d-4)J^2 S^2 + 8(d-2)S^4}dJ^2 \bigg).  
\end{split}
}
\end{widetext}
where 
\be{
%\lambda = \frac{1}{(d-2)(S^2 + 4J^2)\big[(d-3)S^2 + 4(d-5)J^2\big]S}.
%\lambda = \frac {1} {4(d-2) (S^2+4J^2)^{\frac{d+1}{d-2}} S^{\frac{2d-5}{d-2}}}.
\lambda = \frac {1} {4(d-2) (S^2+4J^2)^{\frac{2d-5}{d-2}} S^{\frac{d+1}{d-2}}}.
}
This metric can be brought into a diagonal form via coordinate transformations  
\be{
u = \frac{J}{S}
} 
and 
\be{
\label{eq:tau}
\tau = \sqrt{\frac{d-2} {d-3}}\: S^{\frac{d-3}{2(d-2)}}(1+4u^2)^{\frac{1}{2(d-2)}}. 
}
%Inverting (\ref{eq:tau}) we obtain the entropy of the $d$-dimensional Kerr black hole in the new coordinates as 
%\be{
%S = \paren{\frac {d-3} {d-2}}^{\frac{d-2}{d-3}}\frac{\tau^{\frac{2(d-2)}{d-3}}}{(1+4u^2)^{\frac{1}{(d-3)}}}. 
%}
The Weinhold metric in a diagonal form now reads:
\be{
ds^2_W = - d\tau^2 + {\frac{2(d-3)}{(d-2)}}{\frac{(1- 4{\frac{d-5}{d-3}}u^2)}{(1 +4u^2)^2}}\tau^2 du^2.
}
This metric is a flat metric. In $d=4$ we can write it in Rindler coordinates as
\be{
ds^2_W = -d\tau^2 + \tau^2 d\sigma^2
}
by using
\be{
u = \onehalf \sinh 2\sigma. 
}
In four dimensional spacetime we have the extremal limit along $J/M^2 = 1$ hence $u$ is bounded by 
\be{
\abs{u} \leq \onehalf  \Leftrightarrow \abs{\sigma} \leq \frac{1}{2} \sinh^{-1}1 \approx 0.4406.
}
By using (\ref{eq:Minkowski-coord}) we obtain the wedge of the state space of the Kerr black hole (see FIG. \ref{fig:wedgeKerr}) in a flat Minkowski space whose edge is bounded by 
\be{
-\sqrt{\frac{\sqrt{2}-1}{\sqrt{2} + 1}} \leq \frac{x}{t} \leq \sqrt{\frac{\sqrt{2}-1}{\sqrt{2} + 1}} \:.
}
%-----------------------------------------------------------
%-----------------------------------------------------------
In five dimensions, the extremal limit is given by $J^2/M^3 = 16/27$ and $-\infty \leq u \leq  \infty $ where
\be{
u = \onehalf \tan \sqrt{3}\sigma. 
}
Hence we obtain a wedge with a different opening angle since $\sigma$ falls in the range 
\be{
\abs{\sigma} \leq \frac{1}{\sqrt{3}}\arctan \infty = \frac{\pi}{2\sqrt{3}} \approx 0.9069.
}
The opening angle of the wedge for the $d=5$ Kerr black hole is wider than that of $d=4$. Remarkably, the wedge of the $d \geq 6$ Kerr black hole fills the entire light cone. This is because for black holes in $d \geq 6$ there are no extremal limits.  It is noteworthy that there is a causal structure of state space, but it is determined by curves of constant entropy rather than by the lightcone itself. The curves of constant entropy for $d=4$ Kerr black hole in Minkowskian coordinates are given by 
\be{
S = \frac{(t^2 - x^2)^4}{4(t^2 + x^2)^2}.
}
Differentiation of the mass function (\ref{eq:Kerr-Mass}) with respect to the entropy gives the temperature of the Kerr black hole in arbitrary $d$ as
\be{
\dis
%T = \frac{(d-3)(d-2)^{d-3} M^{3-d}S^{d-4}}{4^{d-2}} \paren{1 + 4\frac{d-5}{d-3}\frac{J^2}{S^2} }. 
%T = \frac{d-3}{4}\paren{1 + \frac{4J^2}{S^2}}\paren{1 + 4\frac{d-5}{d-3}\frac{J^2}{S^2}}S^{-1/(d-2)}.
%T = \frac{(d-3)}{4S^2}\frac{1 + 4\frac{d-5}{d-3}\frac{J^2}{S^2} }{ 1+ 4\frac{J^2}{S^2}}.
\dis 
T = \frac{(d-3)\paren{1 + 4\frac{d-5}{d-3}\frac{J^2}{S^2}}}{4S^\frac{1}{d-2}\paren{1 + 4\frac{J^2}{S^2}}^\frac{d-3}{d-2}}.
}
This temperature shrinks to zero at extremality for $d=4, 5$. According to \cite{myers-perry, ultra-spinning} for $d \geq 6$ there is no extremal limit, the black hole's temperature does not vanish but reaches minimum and starts to behave differently as $T \sim r^{-1}_+$.  In any dimension, we obtain the Ruppeiner metric by using the conformal relation (\ref{eq:conformal}). It is found to be a curved metric with the curvature scalar of the form 
\be{
\dis
\label{eq:Kerr-curvature}
\mathcal{R} =  - \frac{1}{S} \frac{\dis 1 - 12\frac{d-5}{d-3}\frac{J^2}{S^2}}{\dis \paren{1 - 4\frac{d-5}{d-3} \frac{J^2}{S^2}}\paren{1 + 4\frac{d-5}{d-3} \frac{J^2}{S^2}}}.
}
In $d=4$ the curvature scalar diverges along the curve $4J^2 = S^2$ which is consistent with the previous result \cite{ourpaper}.  The Ruppeiner curvature scalar in (\ref{eq:Kerr-curvature}) is valid in any dimension higher than three. In $d=5$ the curvature is reduced to 
\be{
\label{eq:curvature-dim5}
\mathcal{R} = -\frac{1}{S} 
}
which diverges in the extremal limit of the $d=5$ Kerr black hole. For $d \geq 6$ we have a curvature blow-up but not in the limit of extremal black hole, rather at 
\be{
4J^2 = \frac{d-3}{d-5} S^2.
}
This is where Emparan and Myers \cite{ultra-spinning} suggest that the Kerr black hole becomes unstable and changes its behavior to be like a black membrane. Note that in $d=5$ for some values of the parameters, there exist ``black ring'' solutions \cite{emparan-reall} whose entropy is larger than that of the black hole studied in this paper. A careful observation of (\ref{eq:curvature-dim5}) indicates that nothing special happens to the Gibbs surface of the Kerr black hole \emph{\`{a} la} Myers-Perry.

\section{Multiple-spin Kerr black hole}

This is a case where the entropy of the black hole is a function of three parameters, namely the function of mass and two spins. Another example of the three-parameter thermodynamic geometry can be found in \cite{ourpaper} where the Ruppeiner and Weinhold geometries of the Kerr-Newman black hole were investigated. The general Kerr metric in arbitrary dimension $d$ is available in, \eg \cite{myers-perry, ultra-spinning, gibbons-kerr}. The black holes have $(d-1)/2$ angular momenta if $d$ is odd and $(d-2)/2$ if $d$ is even. The multiple-spin Kerr black hole's metric in Boyer-Lindquist coordinates for odd $d$ is given by 
\be{
\begin{split}
ds^2 = &-d\bar{t}^2 + (r^2 + a^2_i)(d\mu^2_i + \mu^2_i d\bar{\phi}^2_i) \\ 
       & +\frac{\mu r^2}{\Pi F}(d\bar{t} + a_i \mu^2_i d\bar{\phi}_i)^2 + \frac{\Pi F}{\Pi - \mu r^2}dr^2, 
\end{split}
}
where 
\be{
d\bar{t} = dt - \frac{\mu r^2}{\Pi - \mu r^2}dr,
}
\be{
d\bar{\phi}_i = d\phi_i + \frac{\Pi}{\Pi - \mu r^2}\frac{a_i}{r^2 + a^2_i}dr,
}
with the constraint 
\be{
\mu^2_i = 1.
}
The functions $\Pi$ and $F$ are defined as follows:
\be{
\label{eq:parameters-multi-Kerr}
\begin{split}
\Pi &= \prod_{i=1}^{(d-1)/2} (r^2 + a^2_i), \\
  F &=   1 - \frac{a^2_i \mu^2_i}{r^2 + a^2_i}.
\end{split}
}
The metric is slightly modified for even $d$ \cite{myers-perry}. The event horizons in the Boyer-Linquist coordinates will occur where $g_{rr} = 1/g_{rr}$ vanishes. They are the largest roots of 
\be{
\Pi - \mu r = 0 \goodspace \text{even $d$} 
}
\be{
\Pi - \mu r^2 = 0 \goodspace \text{odd $d$} .
}
The areas of the event horizon are given by 
\be{
A = \frac{\Omega_{(d-2)}}{r_+} \prod_i (r^2_+ + a^2_i) \goodspace \text{odd $d$},
}
\be{
A = \Omega_{(d-2)} \prod_i (r^2_+ + a^2_i) \goodspace \text{even $d$}.
}
In $d = 5$ there can be only two angular momenta associated with the Kerr black hole, thus the area of the event horizon reads
\be{
A = \frac{2\pi^2}{r_+} (r^2_+ + a_1^2)(r^2_+ + a^2_2). 
}
The temperature of the $d=5$ Kerr black hole with two spins is the Hawking temperature $T = \kappa / 2\pi $ where the surface gravity $\kappa$ is given by
\be{
\kappa = r_+ \paren{\frac{1}{r^2_+ + a_1^2} + \frac{1}{r^2_+ + a_2^2}} - \frac{1}{r_+}.
}
Since there are two angular momenta, hence two angular velocities are associated with this black hole,
\be{
\Omega_{a_1} = \frac{a_1}{r_+^2 + a_1^2}, \goodspace \Omega_{a_2} = \frac{a_2}{r_+^2 + a_2^2}.
}
The first law of thermodynamics for this black hole takes the form \cite{gibbons-kerr}
\be{
dM = T dS + \Omega_{a_1} dJ_{a_1} + \Omega_{a_2} dJ_{a_2}. 
}
The entropy of the $d=5$ Kerr black hole with double spins is given by
\be{
S = \frac{k_B A}{4G} = \frac{k_B}{4G}\frac{2\pi^2}{r_+} (r^2_+ + a_1^2)(r^2_+ + a^2_2).
}                                           
We can choose $k_B$ and $G$ such that $S$ simplifies as
\be{
S = \frac{1}{r_+} (r^2_+ + a_1^2)(r^2_+ + a^2_2),
}
where $r_+$ is the largest root of 
\be{
(r^2 + a_1^2)(r^2 + a_2^2) - \mu r^2 = 0, 
}
where $\mu$ is the ADM mass defined in (\ref{eq:ADMmass}) with $d=5$ and $\dis a_i = 3J_i/2M$. The temperature of the $d=5$ double-spin Kerr black hole reaches zero in the extremal limit which is given by 
\be{
a_1 + a_2 = \sqrt{\mu}
}
or explicitly in terms of mass and the two spins as
\be{
J_1 + J_2 = \frac{4M^{3/2}}{3\sqrt{3}}.
}
Since solving for the entropy function directly is rather complicated, we thus use the same procedure as in the case of the single-spin Kerr black hole and obtain the mass as a function of entropy and two angular momenta as
\be{
\label{eq:mass-multipleKerr}
\dis
M = \frac{3S^{2/3}}{4}\paren{1 + \frac{4J^2_1}{S^2}}^{\frac{1}{3}}\paren{1 + \frac{4J^2_2}{S^2}}^{\frac{1}{3}}.
}

\begin{center}
\begin{table*}
\begin{tabular}{|c|l|l|l|}
\hline\hline
\textbf{Spacetime dimension} & \textbf{Black hole family} & \textbf{Ruppeiner} & \textbf{Weinhold}  \\
\hline
$d=4$   & Kerr   &  Curved   &  Flat    \\ \cline{2-4}
      & RN       &  Flat     &  Curved  \\ \hline 
$d=5$   & Kerr   &  Curved   &  Flat    \\ \cline{2-4}
      & double-spin Kerr & Curved & Curved \\ \cline{2-4} 
      & RN       &  Flat     &  Curved   \\ \hline 
$d=6$   & Kerr   &  Curved   &  Flat    \\ \cline{2-4}
      & RN       &  Flat     &  Curved   \\ \hline 
\text{any} $d$   &  Kerr     &  Curved   &  Flat   \\ \cline{2-4}
      & RN       &  Flat     &  Curved   \\
\hline\hline
\end{tabular}
\caption{Geometry of higher-dimensional black hole thermodynamics.}
\label{table:summary}
\end{table*}
\end{center}

The Hessian of $M$ with respect to the entropy and two angular momenta yields the Weinhold metric, which is found to be curved. The curvature scalar of the Weinhold metric takes the form
\begin{widetext}
\be{
\mathcal{R}_{\small Weinhold} = \frac{16}{3}\frac{S^\frac{2}{3}(S^8 + 3 S^6 J_1^2 + 3 S^6 J_2^2 + 4 S^4 J_1^2 J_2^2 + 64 J_1^4 J_2^4)}  {(S^2 + 4 J_1^2)^\frac{1}{3} (S^2 + 4 J_2^2)^\frac{1}{3}(S^2 - 4 J_1 J_2)^2 (S^2 + 4 J_1 J_2)^2}.
}
\end{widetext}
We next transform it into the Ruppeiner metric via the conformal relation with an inverse temperature as a conformal factor. The temperature of the double-spin Kerr black hole in five dimensions is given by
\be{
%T = \frac{9}{32}\frac{(S^2 - 4J_1 J_2 )(S^2 + 4 J_1 J_2)}{M^2 S^3}. 
T = \frac{1}{2S^{5/3}} \frac{(S^2 + 4J_1 J_2)(S^2 - 4J_1 J_2)}{(S^2 + 4J_1^2)^{2/3}(S^2 + 4J_2^2)^{2/3}}.
}
The Ruppeiner curvature scalar of the double-spin Kerr black hole in five dimensions reads
\begin{widetext}
\be{
\mathcal{R}_{\small Ruppeiner} = - \frac{ S^8 + 20 S^6 J_1^{2} + 20 S^6 J_2^2  + 256 S^4 J_1^2 J_2^2 + 192 J_1^4 J_2^2 S^2 +  192 J_1^2 J_2^4 S^2 -  256 J_1^4 J_2^4 }{ 2S ( S^{2} + 4 J_{1}^{2} ) (S^{2} + 4 J_{2}^{2})( S^{2} -  4 J_{1} J_{2}) (S^{2} + 4 J_{1} J_{2})}.
}
\end{widetext}
Note that both the Weinhold and Ruppeiner curvature scalars are divergent at 
\be{
J_1 J_2  = \frac{S^2}{4}, 
}
which is the extremal limit of the $d=5$ double-spin Kerr black hole. Note also that this curvature scalar does not vanish either in the limit of $J_1 = 0$ or $J_2 = 0$.

\section{Discussion}

The geometry of black hole thermodynamics in higher dimensional spacetime has a pattern similar to the previous study in $d=4$ spacetime. Since microstructures of black holes are unknown, we cannot yet conclude our findings along the same line as those done for the ideal gas \cite{idealgas}. Examination of our examples shows that when a flat thermodynamic curvature arises, it is the Hessian of a function of the form 
\be{
\psi (x, y)  = x^a f\paren{\frac{x}{y}}
}
with $a$  a constant. We have checked that such a function always gives a flat thermodynamic metric, regardless of $a$, and regardless of the function $f$. 

It is worthwhile to observe that inclusion of a cosmological constant leads to curved thermodynamic geometries \cite{ourpaper}; see \cite{carter} for a discussion of the higher dimensional cases.  We note that the calculations of the three-parameter thermodynamic curvature scalars are best achievable by utilization of computer programs for algebraic computations such as CLASSI \cite{jan} and GRTensor \cite{GRtensor} for Maple.

\section{Summary and Outlook}

In this paper, we study thermodynamic geometries of the black hole families and obtain some interesting results, coinciding with our previous findings. We speculate some sort of duality between entropy and mass of the black hole, which is somewhat corresponding to the very distinction between a rotation parameter and an electric charge. 

We summarize our results in the TABLE \ref{table:summary}. Furthermore, we have seen that the Ruppeiner curvature, in all the systems we have studied so far, behaves in a physically very suggestive way. The question why we have this pattern may be answered by the quantum theory of gravity in the future.

\begin{acknowledgments}

We would like to thank Ingemar Bengtsson for encouraging and insightful discussions as well as valuable comments on this manuscript.  N.P. wishes to thank his parents for financial support.

\end{acknowledgments}

{\bf Added Note:} Ref.\ \cite{israel} contains very relevant information about the five-dimensional case.

\end{document}